\documentclass[final]{svjour2}
\usepackage{graphicx}
\usepackage{rotating}
\usepackage{amssymb}
\usepackage{mathptmx}
\usepackage[numbers]{natbib}
\makeatletter
\journalname{}

\bibpunct{}{}{,}{s}{}{,}

\begin{document}

\newcommand{\hdblarrow}{H\makebox[0.9ex][l]{$\downdownarrows$}-}

\title{Quantized vortex state 
in hcp solid $^4$He}

\author{Minoru Kubota}  

\institute{ Institute for solid State Physics, the University of Tokyo,\\ Kashiwanoha 5-1-5, Kashiwa, 277-8581, Japan \\
Tel.:-81-4-7136-3355\\ Fax:-81-4-7136-3355\\
\email{kubota@issp.u-tokyo.ac.jp}
}

\date{\today}

\maketitle

\begin{abstract}
The quantized vortex state appearing  in the recently discovered new states in hcp $^4$He since their 
 discovery\cite{kimchan1,kimchan2} is discussed. Special attention is given to  evidence for  the vortex state as  the vortex fluid (VF) state\cite{VFAnderson,VFAnderson2,VFPenzev,Nemir0907} and its transition into the supersolid (SS) state\cite{VFtoSS,QFS2009Kubota,VFtoSS2011}. 
 Its features are described. The historical explanations\cite{ReattoChester,Chester70,AndreevLifshitz69,Leggett70,matsudatsuneto70} for the SS 
 state in quantum solids such as solid $^4$He 
were based on the idea of Bose Einstein Condensation (BEC) of the imperfections such as vacancies,  interstitials and other possible excitations in the quantum solids which are expected 
because of the large zero-point motions. 
The SS  state was proposed as a new state of matter in which real space ordering of the lattice structure of the solid coexists with the momentum space ordering of 
superfluidity.
%
%
A new type of superconductors, since 
the discovery of the cuprate high $T_c$ superconductors, HTSCs\cite{CuprateSC}, has been 
shown 
to share a feature with 
the vortex state, involving the VF 
 and vortex solid states. 
The high  
$T_c$s of these materials 
 are being discussed in connection to the large fluctuations associated with some other phase transitions like 
the antiferromagnetic transition in addition to that of the low dimensionality. 
%
%
The supersolidity in the hcp solid $^4$He, in contrast 
to the new superconductors which have 
multiple degrees of freedom 
of the Cooper pairs with spin as well as angular momentum freedom, has a unique feature of  possessing possibly only the momentum fluctuations and vortex ring excitations associated with the possible low dimensional fluctuations of the subsystem(s). The high onset temperature of the VF 
state can be understood by considering thermally excited low D quantized vortices and it 
 may be necessary  
  to seek 
  low dimensional sub-systems in hcp He which are hosts for vortices. 
%

\end{abstract}


\section{Introduction}
 \label{sec:1}

Quantized vortices are the essence of superfluidity\cite{SuperFlowDissipationRMP} and superconductivity (superfluidity of electrons). They originate in the coherence of the wave function describing the system over macroscopic length scales much larger than the vortex core sizes. In the $"$classical$"$ superfluids, the macroscopic coherence has been provided by the Bose-Einstein Condensation (BEC) of the constituent Bose particles\cite{SF_BEC}, and that of Cooper pairs\cite{BCS}, in the case of Fermion systems. The superfluid-normal fluid transition has been discussed since the classical discussion by Feynmann\cite{Feynmann} as a phase transition by the destruction of the macroscopic coherence by excitations of quantum vortex loops, and it is discussed much more concretely by a simpler vortex rings model by Williams\cite{WilliamsLambdaTransition}. More recently, 
since the discovery of the cuprate high $T_c$ super-conductors, HTSCs\cite{CuprateSC}, people realized that we do have the quantized vortex state, which is characterized by the complex
$T-H$ phase diagram of the new superconductors, namely, a VF 
 state and different types of vortex solid states have been realized. 
All the $"$new type of superconductors$"$ 
possess a common 
feature, namely the unique vortex state\cite{VS}. It is also true for the newly found FeAs based superconductors\cite{FeBasedSC}. Early discussions for the vortex state in the cuprates argued for the importance of the 2D character of the CuO$_2$ plane Cooper pairs for the cuprate HTSCs\cite{FFH}. This kind of appearance of the vortex state in the new superconductors, the strongly correlated electron systems, is often discussed nowadays in connection to the 2D pancake-type vortices\cite{clem97}, but is never discussed for any mass superfluidity nor for the physics of solid He with a single exception\cite{AndersonSupersolid}, though solid He is actually a candidate for being one of the most strongly correlated systems.

Ideal 2D superfluidity and the physics of 2D vortices had been discussed for the superfluidity of the 2D $^4$He films by Berezinskii\cite{B}, Kosterlitz and Thouless\cite{KT} and lead to the idea of topological BKT phase transition, where real BEC is absent, but 2D quantized vortices and local condensate were implicitly supposed. Its unique features, the 2D density linear transition temperature $T_{KT}$ and the universal jump\cite{UnivJump} in the superfluid density are clearly experimentally confirmed\cite{bishopreppy78,bishopreppy80}. A detailed discussion of the dynamics of the KT systems was given by Ambegaokar, Halperin,  Nelson, and Siggia\cite{AHNS}.
KT superfluidity of the CuO$_2$ plane electron Cooper pairs  and Josephson coupling between the layers are the essence of the HTSC vortex physics. A unique vortex state, first discussed for the cuprate 
HTSCs 
 supposes  thermal excitations\cite{FFH} of low dimensional quantized vortices, which have low enough energy and high entropy in high fields.  

Readers are recommended to refer to a longer paper by the present author on the vortex physics in hcp $^4$He (I), which includes a discussion on the essence of superfluidity and 
the vortex physics in 3D He film systems, where detection of 3D vortex lines penetration  by torsional oscillator (TO) technique is also described\cite{KubotaQVPhys}.

\section{"Possible supersolidity" 
and 
Quantized vortex physics  in hcp $^4$He}
 \label{sec:2}

The discussion of  possible superfluidity in the quantum solid, the SS 
 state, had  historically assumed a non-negligible amount of vacancies, interstitials, and other possible excitations in the quantum solid\cite{Chester70,AndreevLifshitz69,Leggett70}, because of the  zero point motions in such systems and the possible Bose Einstein condensation BEC of these quasi particles or excitations.  
It was believed to certainly occur at some sufficiently low  temperature. In 1970  no other possible 
mechanism for superfluidity in the solid was  known. The BKT mechanism \cite{KT,UnivJump,bishopreppy78,AHNS}  for a  2D system was first introduced in 1972 and the lack of 
knowledge of other mechanisms like those possible in 1 D 
systems 
 lead to this situation. It was later when a 
1D dislocation network was considered as a candidate explanation for a possible SS 
 state\cite{shev87,shev88}. 

New  development started when Kim and Chan reported their new experimental results using torsional oscillator TO 
technique, \cite{kimchan1,kimchan2} which raised the question whether or not they finally
 found the long sought, non-classical rotational inertia, $NCRI$ \cite{Leggett70}  of the SS 
  state.
 There have been frequent international workshops since 2006 when other groups started confirming results, yet the system remains 
 still mysterious even after so many years of work by an increasing number of groups. 
By now several 
 reviews\cite{Prokfev,GalliReattoR,SSDisorder,Balibar2010}  have appeared since the restart by Kim and Chan\cite{kimchan1,kimchan2}. 
The main problem is that nobody knows what a 
SS 
 state should look like. There are plenty of interesting aspects of the solid as well in the quantum solid He. First of all, Day and Beamish\cite{ShearModulus} reported an increase of the shear modulus towards lower temperatures near 
200 mK, almost the same temperature range as TO anomalies had been reported. Substantial change of the properties by a minute amount of $^3$He 
 is also reported since the beginning
\cite{kimchan1}. 
A more detailed discussion of these peculiar behaviors is left for 
later 
in the discussion chapter.
In the present chapter discussion is 
concentrated on the problem of whether quantized vortices are present in the system or not and 
 if there is enough evidence supporting supersolidity and related phenomena in solid He.

Our discussion 
started with solid He, but now we limit ourselves to 
the hcp phase of the solid $^4$He, because of the following development: 
The  shear modulus increases 
below around 200 mK for hcp $^4$He\cite{ShearModulus}. 
A similar 
 change 
was also found\cite{statistics} in hcp $^3$He, 
 but the  TO response was found only in hcp $^4$He.  

 \begin{figure}
  \center
  \includegraphics[width=0.45\textwidth]{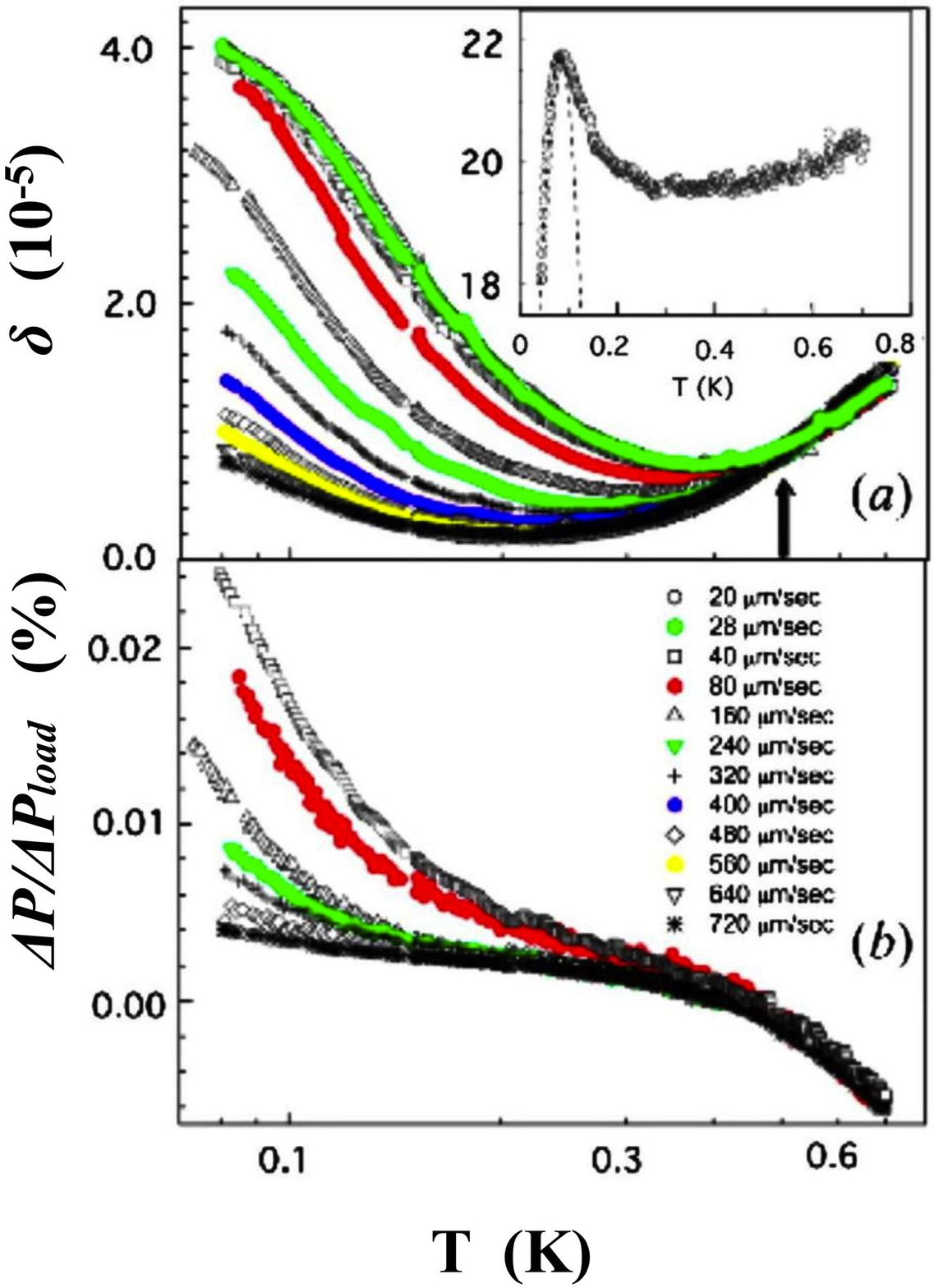}
\caption{$T$ dependence of reduced energy dissipation $\delta$(a) and  $\Delta$$p$$/$$\Delta$$p_{load}$(b) 
at various edge velocities 
 $V_{ac}$ in $\mu$m/s for a 32 bar sample\cite{VFPenzev}. The values of $\delta$ are presented without any artificial shift of data, which has often been needed in other papers. Careful treatment of the cryostat is essential. 
 Some data are omitted for clarity [all of the data on $V_{ac}$ dependence are plotted in Fig.2 of Penzev $\it {et}$  $\it {al.}$\cite{VFPenzev}(a) and (b)]. An arrow in (a) indicates $T_o$, across which $V_{ac}$ dependence changes. The inset in (a) indicates a typical dissipation peak with 
 higher $T_p$. The low $T$ part of the peak was fitted with a Gaussian function. 
  The zero of 
  $\Delta$p/$\Delta$$p_{load}$ in (b) is taken 
 where $V_{ac}$ dependence disappears 
  near $T_o$.
}
\label{onset}  
\end{figure}

\subsection{Quantum Vortex State in hcp $^4$He}
 \label{sec:2.1}

Although enough details of the microscopic mechanism responsible for what is occurring in  hcp $^4$He are still not available to explain all the all the experimental observations, the existence of quantized circulations or quantized vortices can be shown by experimental evidence and it should be regarded as fundamental to determine the direction of the research activities. 
%
In this chapter 
 experimental discoveries are discussed which are supporting quantized vortices in the VF 
 state in hcp solid $^4$He, a unique state 
 among all the known superfluids. It is, however, quite common to discuss the VF state\cite{VFAnderson,VFAnderson2} in the so called  $"$new type of  superconductors$"$\cite{aboveTc,QL}, though there have not been many microscopic arguments from the standpoint of quantized vortices in the new type of superconductors.  Anderson\cite{AndersonSupersolid} has recently proposed an explanation which may cover a large portion of various experimental observations starting from the VF state picture.


The next subsection of this chapter starts with  experimental discoveries of the peculiar TO drive velocity $V_{ac}$ dependence observed in hcp solid $^4$He (Fig.~\ref{onset}), which turned out to indicate the unusual behavior of the VF state and that of onset of the VF state. Subsection \ref{sec:2.3} deals with vortex dynamics in the VF state, utilizing the standard theoretical method of handling the tangled vortices in  superfluid turbulence\cite{Nemir0907}. In chapter~\ref{sec:3} 
 the recent discovery of the transition from the VF state into the real SS 
 state\cite{VFtoSS,QFS2009Kubota,VFtoSS2011} is reported and 
  its  properties are discussed in terms of 
the SS 
 density $\rho_{ss}$ and the critical velocity $V_c$.


\subsection{Discovery of the onset temperature $T_o$ of the Vortex Fluid(VF) State  in hcp  $^4$He and its unique properties
}
\label{sec:2.2}


Actually the 
vortex phase diagram of hcp  $^4$He was first experimentally proposed by reporting a definite onset temperature by Penzev $\it {et }$ $\it {al.}$\cite{VFPenzev}, 
after Anderson's suggestion\cite{VFAnderson,VFAnderson2} and by describing unique features of the VF state, which appeared to be characterized by thermally excited vortex fluctuations\cite{Nemir0907}.  These thermally excited fluctuations  cause the unique drive amplitude dependence of the TO response\cite{VFPenzev}. 
Fig~\ref{onset} shows a detailed study of hcp  $^4$He at $T$ above the dissipation peak at $T_p$. The 
$V_{ac}$ at the outer edge of  the cylindrical sample  influences both Period $P$ shift $\Delta$$P$/$P_{load}$ and energy dissipation in the sample $\delta$ quite a lot, but only  below an "onset temperature" $T_o$. As noted before, although this strong $V_{ac}$ dependence was reported even in the original work by Kim and Chan\cite{kimchan1} nobody discussed quantitatively its dependence, but refer to it as a sort of "critical velocity" phenomenon. 
This speed
 , on the order of 10$\mu$m/s, needs to be discussed as a characteristic speed\cite{VFPenzev} 
 to describe system size fluctuations instead of a $"$critical velocity$"$ in the later section on vortex dynamics in the VF state. The VF state 
  may  not have a critical velocity.
In the "onset" paper\cite{VFPenzev} the authors discussed 
$T_o$ as the onset temperature 
 of the VF state, below which quantized vortices are thermally excited and that $NLRS$ exhibits a unique temperature dependence, which is different from any of the order parameters below a phase transition. Furthermore, they observed rather clear $log(V_{ac})$ linear behavior\cite{highp,VFPenzev} (see also Fig~\ref{Kubota_SS_3}) 
over a decade of $V_{ac}$ for 
the entire $T$ range well above the dissipation peak at $T_p$. 
This dependence is argued to be responsible either for appearance of the quantized vortex lines penetrating 
 the sample, or polarizing  the vortex rings in the VF state, in both cases, 
 thereby reducing the disorder of the thermally excited VF state. This observation is  unique compared with all other TO or other oscillator experimental results for 
known superfluid transitions, KT transition of 2D films\cite{bishopreppy78,bishopreppy80}, bulk He liquids in Vycor\cite{ReppyTylar}, and also in the artificial 3D superfluid made of  $^4$He monolayer superfluid films\cite{FukudaPhD,VLinesThru,KubotaQVPhys}, where energy dissipation $\Delta$$Q^{-1}$ is larger when the oscillation amplitude is driven larger. Their intuitive discussion is also proven in terms of vortex dynamics in the randomly fluctuating vortex tangle\cite{Nemir0907}. 
%
%
%
%
%
The  transition into the SS state\cite{VFtoSS,QFS2009Kubota} will be discussed, which is characterized by a unique order parameter and the critical velocity in the next chapter 3.
\begin{figure}
  \center
\includegraphics[width=0.68\textwidth]{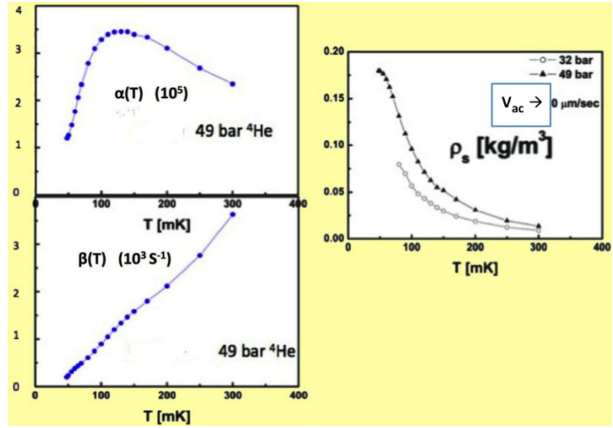}
\caption{Parameters $\alpha(T)$, $\beta(T)$, and $\rho_{s}(T)$= $NLRS$($V_{ac}$$\to$0) obtained from the
data of Fig. 1 of Nemirovskii $\it {et}$  $\it {al.}$\cite{Nemir0907} with the use of analysis described in text. $\beta(T)$ goes to zero, in other words, it means the relaxation time goes to infinity at extrapolated $T$ $\approxeq$ 30 mK. This may imply the possible disappearance of the thermal excitations and 
of the VF 
state.
}
\label{Nem2}  
\end{figure}


\subsection{Vortex dynamics in the VF state and possible disappearance  of the thermal excitations}
\label{sec:2.3}

The VF 
 state has been discussed with regard to 
the H-T phase diagram of the new type of superconductors since discovery of the cuprate HTSCs, yet, there has not been a quantitative discussion from the quantized vortex dynamics viewpoint. There are, however, arguments developed for the 
 vortex dynamics in the superfluid turbulence, which  often deals with physics at $T$=0K, where no thermal excitations exist.  The superfluid turbulence state is excited by various methods, but it is excited by some external flow velocity exceeding a characteristic  critical velocity. The VF
 state, on the other hand, is excited by thermal energy and it is considered that the thermal energy can exceed the critical velocity of the turbulent state for the vortex tangle in the VF state. As has 
 been  seen in the  
 previous section, a peculiar feature of the VF state of hcp  He is that some fluctuations can be reduced by the TO drive excitations\cite{VFPenzev}, namely larger drive velocity reduces the fluctuating signal more than smaller drive excitations, as can be seen in Fig~\ref{onset} and Fig~\ref{VFSS2}, for example.  This is actively studied and there are references for it. 
 Nemirovskii $\it {et}$  $\it {al.}$\cite{QFS2009Kubota,Nemir0907}, consider 
 quantized vortex element dynamics for the experimental data analysis of the TO responses of the VF state in hcp $^4$He, using a phenomenological relaxation model;


Angular momentum of the superfluid fraction appears only due to the presence of either aligned vortices (vortex array) or 
the polarized vortex tangle having nonzero total average polarization
$P$ =$L$$<$s$\textquoteright$ ($\xi$)$>$ along the applied angular velocity $\Omega$ (axis z, the magnitude of $\Omega$  = $V_{ac}$/$R$, where $R$ is radius of the sample). Here L is the vortex line density (total length per unit volume), s($\xi$) is the vector line position as a function of label variable $\xi$ and  s$\textquoteright$($\xi$) is the tangent vector. In the steady-state case there is a strictly fixed relation between the total polarization $L$$<$s$\textquoteright$($\xi$)$>$ and applied angular velocity $\Omega$,\\

\qquad  \qquad 	$\Omega$ = $\kappa$$P$/2 = $\kappa$$L$$<$s'($\xi$)$>$/2.   \qquad   	\qquad 	 $\qquad$  \qquad 	(2.1)
\\

They\cite{QFS2009Kubota,Nemir0907} pointed out that there are two possible mechanisms for relaxation-like polarization of the VF.  
The first is alignment of elements of the vortex lines due to interaction with the normal component (see Tsubota  
 $\it {et}$ $\it {al.}$(2004)\cite{tsubota2004} for detailed explanation). This interaction (mutual friction)
is proportional to the local normal velocity, which in turn is proportional to the rim velocity $V_{ac}$. Thus it
is natural to suppose that polarization $P$ of the vortex tangle due to alignment of filaments along $\Omega$($t$) occurs with typical inverse time $\tau^{-1}$($V_{ac}$) which is proportional to the rim velocity $V_{ac}$.
Let us illustrate according to Nemirovskii $\it {et}$ $\it {al.}$, 
with consideration 
of  
Tsubota $\it {et}$ $\it {al.}$(2004)\cite{tsubota2004}. In the presence of mutual friction there is a torque acting on the line and the angle $\phi$  between axis $z$ and the line element changes according to the equation d$\phi$/dt = $\alpha$($V_{ac} /R$) sin$\phi$
 ($\alpha$
%
 is the friction coefficient, dependent, in general, on $T$ and pressure $p$). Except for a short transient, the solution to this equation can be described as a pure exponential
$\thicksim$ exp(-t/$\tau_1$($V_{ac}$)), with the velocity dependent inverse
time $\tau_1$$^{-1}$($V_{ac}$) $\thicksim$ $\alpha$($V_{ac}$)/$R$. 
Thus, it is concluded that during time-varying rotation or torsional oscillation vortex filaments tend to align along the angular velocity direction. However, there can be not enough pre-existing vortex lines in the tangle to involve all the superfluid part into the rotation to satisfy the relation (2.1), or on the contrary the initial vortex tangle can be excessively dense.

In such a case deficit (extra) vortices should penetrate into (leave from) the bulk of the sample. This 
penetration occurs in a diffusion-like manner\cite{nemir0902} and leads to the relaxation-like saturation of the vortex line density $L$(t) Nemirovskii $\it {et}$  $\it {al.}$\cite{QFS2009Kubota,Nemir0907} 
 assume that this saturation occurs in an exponential manner with some characteristic inverse time  $\tau_2^{-1}$ = $\beta$. Due to linearity of the diffusion process it is supposed that the coefficient $\beta$ is velocity independent, but can be a function of $T$ and $P$. 
Combining these two mechanisms it is assumed that the total polarization of the VF 
 occurs in a relaxation-like manner with a pure exponential function  $\phi$($t$$'$/$\tau$) $\sim$ exp($t$$'$/$\tau$). The inverse time $\tau^{-1}$ of relaxation is just the sum of $\tau_1^{-1}$($V_{ac}$) and $\tau_2^{-1}$: \\

\qquad  \qquad 	$\tau^{-1}$ = $\alpha$($T$)$V_{ac}$/$R$+ $\beta$($T$). \qquad  \qquad   \qquad \qquad  (2.2)
\\

The parameters $\alpha$, $\beta$, and $\rho_s$($T$)$\equiv$$NLRS_0$$(T)$=$\Delta$$P/$$\Delta$$P_{load}$ at $V_{ac}$$\to$0,  obtained using experimental data analysis for a 49 bar hcp $^4$He sample, are given in Fig.~\ref{Nem2}. The 
 $\Delta$$Q^{-1}$and 
 $NLRS$ at different $T$ could be well reconstructed from  these parameters and plotted in Fig. 3 of Nemirovskii $\it {et}$  $\it {al.}$\cite{Nemir0907}.  $\rho_s$($T$) can be regarded as the VF state SS 
  density  and it will be contrasted with the real SS 
   density $\rho_{ss}$($T$) which will be discussed in chapter 3.

Now let us consider what all these results of the measurements as well as the data analysis mean. An interesting observation of Fig.~\ref{Nem2}  $\alpha$ and especially $\beta$  is that they are approaching  zero when $T$ approaches about 30 mK while $\rho_s$ is gradually increasing towards lower temperatures. That means the relaxation time $\tau$ is becoming infinite 
and vortices are becoming difficult to move, while 
 $\rho_s$ is high. This situation reminds us of the dissipation peak appearing near the KT transition\cite{AHNS}, where it is interpreted that the thermally excited 2D vortices loose 
 number density rapidly and  the dissipation peak is localized near the transition at $T_{KT}$, while the superfluid density remain towards $T$=0. The VF state may have lost the thermal activation 
 below $T$=$\thicksim$30 mK.

 \begin{figure}
  \center
  \includegraphics[width=0.55\textwidth]{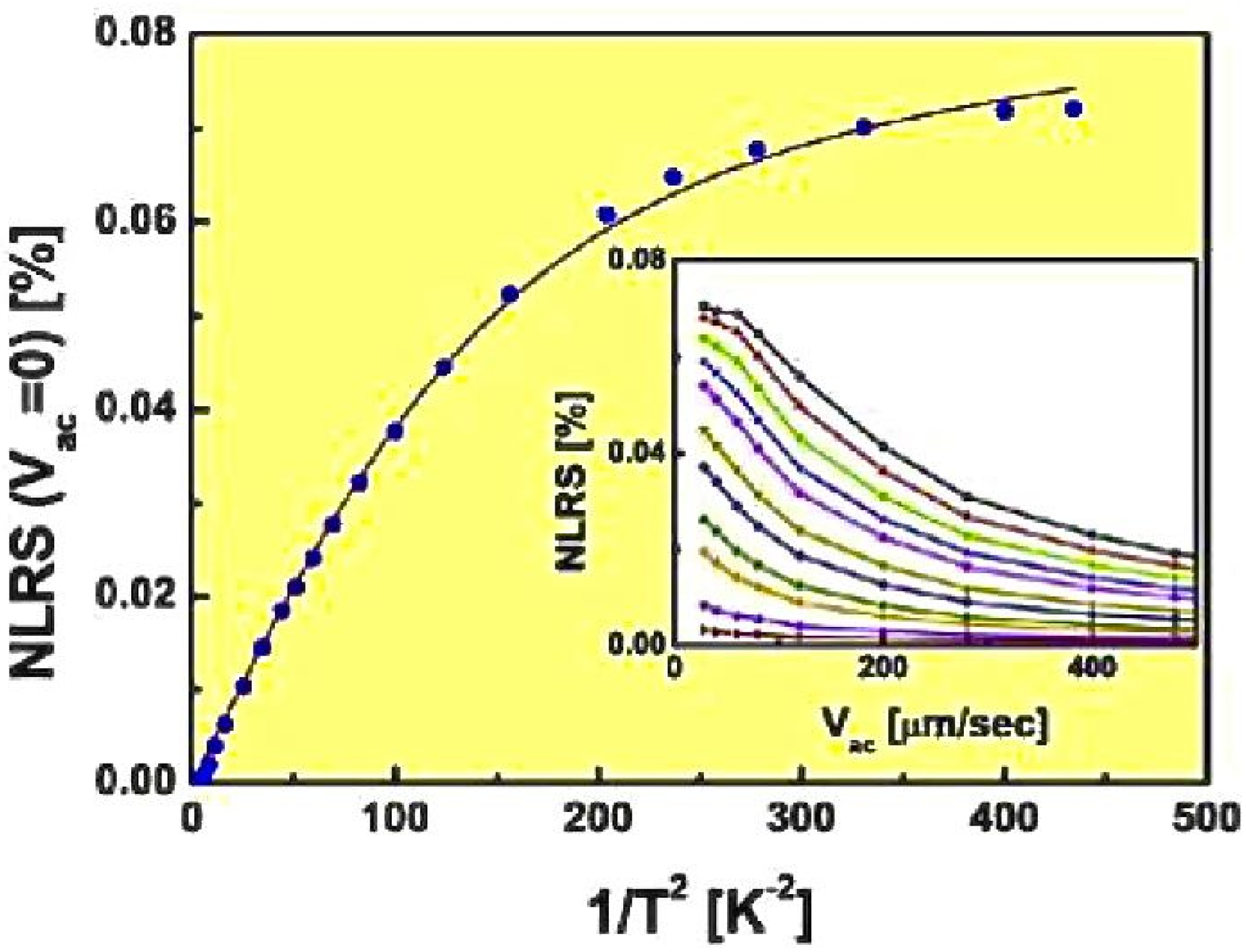}
  \caption{$NLRS_0$$(T)$=$\Delta$$P/$$\Delta$$P_{load}$ at $V_{ac}$$\to$0, (It may be called $\rho_s$ of the VF state to distinguish it from the supersolid density  $\rho_{ss}$) is displayed\cite{VFtoSS} as a function of $x =1/T^2$. The solid line through data points is the Langevin function
  $f(x)$ = a$\lbrack$coth(bx) -1/(bx)$\rbrack$  
    with a = 0.0878 $\pm$ 0.0011, and b = 0.0148$\pm$ 0.0004. The inset shows the $V_{ac}$ dependence of the data at ten temperatures below 300 mK, for clarity on linear scales. It can be safely extrapolated to $V_{ac}$$\to$0. See for details Fig.2 of Penzev $\it {et}$ $\it {al.}$\cite{VFPenzev}.
}
\label{VFSS1}   
\end{figure}

Another interesting observation for the VF 
 state of hcp $^4$He is expressed in Fig.~\ref{VFSS1}. 
 The $logV_{ac}$ linear behavior at larger $V_{ac}$ values than about 40$\mu$m/s linear velocity has been already discussed, and now let us look at the situation when $V_{ac}$ approaches zero. 
The line through the data points in Fig.~\ref{VFSS1} is the Langevin function with $x = 1/T^2$, 
 $f(x)$ = a$\lbrack$coth(bx) -1/(bx)$\rbrack$  
  with a = 0.0878$\pm$0.0011, and b = 0.0148$\pm$0.0004. 
 Various other possible fits were tried, but they failed, for example, with $x =1/T$.   
 The origin of this interesting temperature dependence is 
  not yet 
   known.  Actually this $1/T^2$ dependence appears repeatedly in the following analysis as well.
The Langevin function with $x=1/T$ is followed by the generalized susceptibility of an ensemble of classical dipoles with fixed number.  It is not clear if  the good fit  
 is just accidental or not. 

If one were to consider the 3D counterpart\cite{VFAnderson,VFAnderson2} of BKT theory\cite{B,KT}, one would need to consider the behavior of the ensemble of vortex rings, which are thermally excited in local domains, and a transition into the SS 
 state
would occur because of the coherence length growth towards zero temperature of the low dimensional subsystems, just 
as in the 3D superfluid made of He films
\cite{FukudaPhD,VLinesThru}.

\section{Discovery of the transition into the SS state from the VF 
 state in hcp $^4$He: 
Hysteresis found below a fixed temperature, $T_c$}
\label{sec:3}

 \begin{figure}
  \center
  \includegraphics[width=0.38\textwidth]{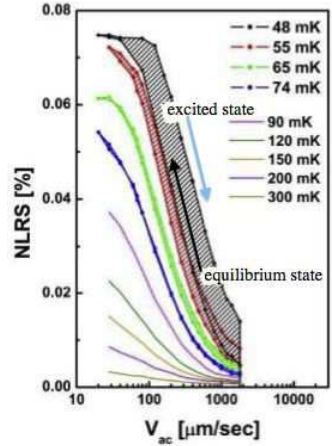}
\caption{The $log(V_{ac})$ dependence of $NLRS$=$\Delta$$P$/$\Delta$$P_{load}$ of our solid $^4$He sample at 49 bar at constant temperatures as given in the figure. Measurements were performed at each $T$ starting at the largest drive which corresponds to 1,800$\mu$m/s and then to lower drive velocity, stepwise. After reaching the minimum drive velocity measurement at $\sim$27$\mu$m/s and after observing  equilibrium,  then upwards at each $T$ for the remaining measurements.  
 Appearance of the hysteretic behavior is found starting at 74 mK and lower temperatures. Limited 
data are shown for clarity. 
}
\label{VFSS2}       
\end{figure}

When Anderson proposed the vortex fluid state to explain 
the reported experimental observation until that time, he pointed out that the real SS 
 state transition should be located at much lower temperature\cite{VFAnderson,VFAnderson2}.

The occurrence of the unique hysteresis in the TO experiments with solid $^4$He samples when the excitation velocity $V_{ac}$ was changed at low temperature had been reported first by Kojima's group,\cite{hysKojima} and then by Chan's group,\cite{hysChan}, and then by Reppy's group.\cite{hysRep}  They could not determine 
that the hysteresis is connected with a definite transition at a fixed temperature.  Shimizu $\it {et}$ $\it {al.}$, \cite{VFtoSS} 
found for the first time that the hysteresis occurs
 below a characteristic temperature $T_c$  in solid He. They\cite{QFS2009Kubota,VFtoSS,VFtoSS2011}  propose from their detailed study of the hysteretic behavior that the hysteretic component of the $NLRS$, $NLRS_{hys}$ is actually additive to the $NLRS$ quantity of the VF state.  $NLRS$ is found to  continue to grow towards lower temperatures from the onset temperature $T_o$ as shown in Fig.~\ref{onset}. It has a unique, continuous temperature dependence towards $T$ = 0 K as depicted in Fig.~\ref{VFSS1}.
 The additive behavior of the hysteretic component can be also seen in Fig.~\ref{VFSS2}.
. 
The expected SS state should be characterized by a definite order parameter, supersolid density $\rho_{ss}$, in addition to a critical velocity $v_c$ to destroy this state. 
 These points  are discussed in much more detail  in the following section.  

\begin{figure}
\begin{center}
  \includegraphics[width=0.38\textwidth]{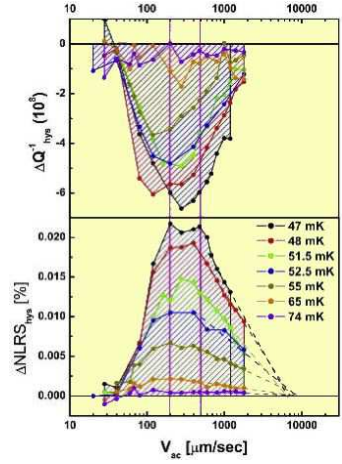}
\end{center}
\caption{The hysteretic components of the TO responses,  $\Delta$$Q^{-1}_{hys}$  as well as $\Delta$$NLRS_{hys}$ $=$ $NCRIF$ are plotted against  $logV_{ac}$. The
hysteretic behavior starts at $V_{ac}$$\sim$40$\mu$m/s at all temperatures below $\sim$75 mK. Both $\Delta$$NLRS_{hys}$	and	$\Delta$$Q^{-1}$	reach	some	extreme	values	for	the	range $V_{ac}$  $=$ 200$-$500$\mu$m/s. Then the absolute size of both quantities starts
to decrease. 
 Nearly $log$$V_{ac}$ linear decreases were observed, especially for $\Delta$$NLRS_{hys}$. From this straight 
 extension of the $logV_{ac}$ linear dependence, 
  a critical velocity, $\sim$ 1 cm$/$s,  to suppress the $NCRIF$ $=$ $\rho_{ss}$ to zero, was obtained, which compares well within an order of magnitude with $V_c$ $=$ $h$$/$($m_4$$\xi_0$$\pi$) $=$ 6$-$12 cm$/$s, for $\xi_0$ $=$ 25 $-$ 50 nm; see text.}
\label{VFSS3}       
\end{figure}

\subsection{Hysteretic component of the TO responses and possible  SS 
density}
\label{sec:3.1}

As pointed out above, the hysteretic components of the TO response are found to be additive to the response of the VF state.  See Fig.~\ref{VFtoSS2}.
Shimizu $\it {et}$ $\it {al.}$,\cite{VFtoSS} as well as Kubota $\it {et}$ $\it {al.}$,\cite{QFS2009Kubota,VFtoSS2011}  
 tried to analyze the hysteretic components separately from those for the VF state. Fig,~\ref{VFSS3} shows the dissipation change $\Delta$$Q^{-1}_{hys}$(upper column) and the $\Delta$$NLRS_{hys}$ change across  the hysterisis vs $logV_{ac}$(lower column).   
 The lower branch of the $NLRS_{hys}$ is named as the "equilibrium" state and the upper branch as the "excited" state branch.  Fig,~\ref{VFSS3} shows the difference between the "excited" state and "equilibrium" state.  The actual experimental method is described in detail in Kubota $\it {et}$ $\it {al.}$,\cite{VFtoSS2011}. Measurements are performed at a fixed temperature and after equilibrium is established first  at high $V_{ac}$ 
 of the TO and then going 
to lower excitations, step wise after reaching $"$equilibrium$"$ or steady state at each step, to the lowest AC velocity on the order of 8 to 30$\mu$m/s.Then $V_{ac}$ is  increased,  also stepwise, up to the maximum speed 1,800$\mu$m/s.

\begin{figure}
 \center
  \includegraphics[width=0.5\textwidth]{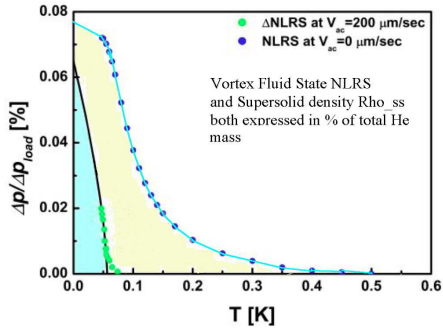}
\caption{ $NLRS$=$\Delta$$P$/$\Delta$$P_{load}$ of the vortex fluid (VF) state expressed in $\%$ of                    
 He mass\cite{VFtoSS,QFS2009Kubota,VFtoSS2011} and the hysteretic component of $NLRS$, $\Delta$$NLRS_{hys}$, which is assumed to be the SS 
  density $\rho_{ss}$, as a
function of $T$ (for details\cite{VFtoSS2011}). At the high temperature end, $NLRS$ behaves as $1/T^2$, whereas $\rho_{ss}$ starts below $T_c$ and increases towards $T$= 0. The latter behaves as an order parameter unlike 
 the $NLRS$. The behavior of the $NLRS$ in the VF looks 
'paramagnetic', with a Langevin function modified with $x = 1/T^2$. Langevin function behavior is expected for the susceptibility of an ensemble of classical dipoles, usually with $x =1/T$. Arranged from Fig.2 of Kubota $\it {et}$ $\it {al.}$\cite{QFS2009Kubota}.
}
\label{VFSS4}       
\end{figure}

What is observed in Fig,~\ref{VFSS3} is a characteristic $V_{ac}$ value of about 40$\mu$m/s, above which hysteretic components develop. $NLRS_{hys}$ reaches a maximum value at 200 $\mu$m/s and then stays almost constant until about 500$\mu$m/s and then it decreases almost linearly in the semi log plot as in Fig,~\ref{VFSS3}.  With linear extrapolation in the semi log plot of the figure, it is observed that the $NLRS_{hys}$ component disappears at $V_{ac}$ about 10,000$\mu$m/s or 1 cm/s. 
 A critical velocity $V_c$ may have been found to destroy the $NLRS_{hys}$  component.
While $NLRS$ increases when it is converted to the excited state,  $\Delta$$Q^{-1}_{hys}$ shows negative changes. This negative change indicates  dissipation in the "excited" state is less than in the "equilibrium" state and implies that vortices are expelled from the sample. 
Shimizu $\it {et}$ $\it {al.}$\cite{VFtoSS}, as well as Kubota $\it {et}$ $\it {al.}$\cite{QFS2009Kubota,VFtoSS2011}, 
plotted the maximum of this $NLRS_{hys}$(at $V_{ac}$=200$\mu$m/s) 
as a function of temperature together with the $NLRS$ extrapolated to $V_{ac}$ = 0 in Fig.~\ref{VFSS4}. While $NLRS$ of the VF state gradually appears as $1/T^2$ below $T_o$ $\sim$500 mK, the hysteretic component $NLRS_{hys}$(at $V_{ac}$=200$\mu$m/s) starts to appear much more sharply below the temperature $T_c$ =75 mK and then changes  slope around 60 mK.

\begin{figure}
 \center  
  \includegraphics[width=0.5\textwidth]{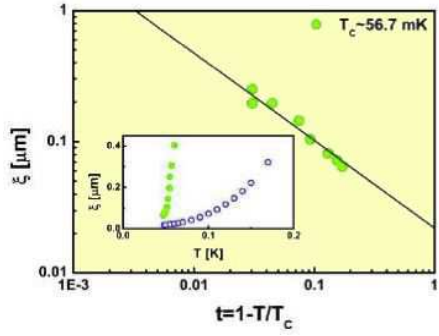}
\caption{Critical behavior of $\xi$ obtained from the data in Fig.~\ref{VFSS2}. 
 $T_c$= 56.7 mK was chosen, supposing $\rho_{ss}$= (1-$T/T_c$)$^\gamma$ = $t^\gamma$ with $\gamma$=2/3. 
  $\xi_0$$\sim$25 - 50 nm was obtained by simple extrapolations, horizontal and straight extension 
   to $t$=1 or $T$=0K. Inset shows $\xi$ for the SS, $\xi_{SS}$ , as well as for the VF state,	$\xi_{VF}$ (open circles
  ),	with	linear	scales.	$\xi_{SS}$ $>$ $\xi_{VF}$ 	at	all	$T$	where	they coexist.
}
\label{VFSS5}       
\end{figure}

Furthermore, they\cite{VFtoSS,VFtoSS2011} checked this $NLRS_{hys}$(at $V_{ac}$=200$\mu$m/s) in various ways and found an interesting feature that they could fit as  (1-$T/T_c$)$^\gamma$ with $\gamma$=2/3 by choosing $T_c$=56.7 mK. This is the critical behavior  expected for  a 3D superfluid transition described with 3D XY model. Using the Josephson's length relation\cite{Josephsonslength1,Josephsonslength2,guzai1} $\xi$ = $\kappa$ $k_B$$T_c$$/$$\rho_s$, one can evaluate the coherence length for the SS state as plotted in Fig.~\ref{VFSS5}, 
%
%
for the hcp He 49 bar sample 
as a function of reduced temperature, $1-T/T_c$ with $T_c$=56.7 mK. The inset of the figure compares the Josephson's coherence length $\xi$ for the VF state $NLRS$ and for the SS state 
  $NLRS_{hys}$(at $V_{ac}$=200$\mu$m/s).
It is seen that the coherence length evaluated in the same manner differs 
 for the VF state and for the SS state. The former is shorter for the same temperature. This may give 
 some clue for the 
 relation between the VF and the SS states, namely the SS state develops a macroscopic scale coherence, while the length of the VF state coherence is 
  limited. 
From all of these trials, it is proposed that $NLRS_{hys}$(at $V_{ac}$=200$\mu$m/s) is actually the real SS 
 density $\rho_{ss}$. 
 This proposal will be considered later once again in connection with vortex lines penetration phenomena under DC rotation. 
In connection to the above question of the 
relation between the VF state and the SS states and the transition between them, 
 the details of the VF state $NLRS$ behavior as a function of 
 $V_{ac}$ will be checked once again in the next section.

\subsection{Further detailed  study of the $logV_{ac}$ linear dependence study
 of the VF state $NLRS$ and transition into the SS 
  state } 
\label{sec:3.2}

\begin{figure}
 \center
  \includegraphics[width=0.40\textwidth]{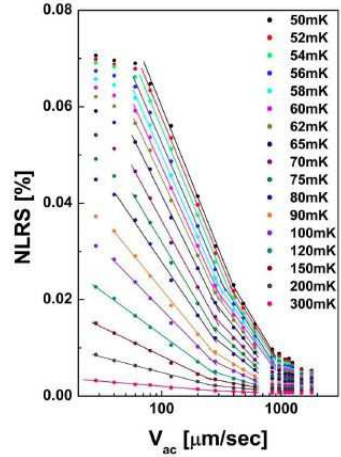}
\caption{Detailed data of 
 $NLRS$ of
 hcp $^4$He 49 bar sample, as a function of $logV_{ac}$,
 \cite{VFtoSS,VFtoSS2011}.
It is observed that
 the $logV_{ac}$ linear relation is followed over
some range of $V_{ac}$, starting at a certain $V_{ac}$ value. Then some other $logV_{ac}$  linear like dependence appears at a higher $V_{ac}$ range for each of data at a $T$. Expanded details are discussed by Kubota $\it {et}$ $\it {al.}$ \cite{VFtoSS2011}, in its Fig.4.and 5. The $logV_{ac}$ linear relation was discussed for solid $^4$He first by Anderson\cite{VFAnderson,VFAnderson2} as evidence
of quantized vortex lines involvement in the physics of solid $^4$He, and then was experimentally demonstrated
 by Penzev $\it {et}$ $\it {al.}$\cite{VFPenzev} for  
 ÓpolarizingÓ a tangled state of the
VF state.
}
\label{Kubota_SS_3}       
\end{figure}

\begin{figure}
 \center
  \includegraphics[width=0.5\textwidth]{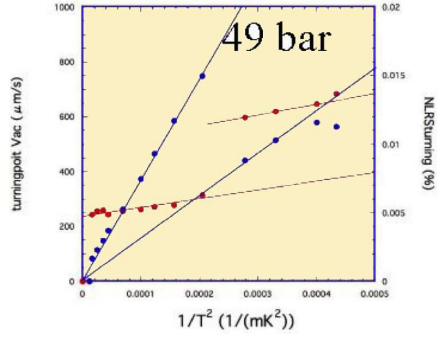}
\caption{The turning point values of $NLRS$ as well as those of the 
velocity $V_{ac}$ are plotted vs $1/T^2$. 
The turning point seems to show considerable 
 regularity. 
A transition was found at $1/T^2$ $=$ 0.00025 $\pm$ 0.00004 $K^2$, or  $T$ $=$ 59$-$69 mK,
which coincides with other $T_c$ determinations. 
}
\label{Kubota_SS_4} 
\end{figure}

Having seen 
 that the hysteretric behavior starts at a finite temperature and the excited state shows a higher value of the $NLRS$ than the equilibrium state, which seems to be a continuation of the VF state from higher temperatures to  lower temperatures, it is questioned 
 if the VF state itself may undergo 
some change at $T_c$. Kubota $\it {et}$ $\it {al.}$\cite{VFtoSS2011} studied in detail the 
$V_{ac}$ dependence of the $NLRS$ in the $"$equilibrium$"$ state. Fig.~\ref{Kubota_SS_3} shows a  $logV_{ac}$ linear relation over a decade of $V_{ac}$ value for each of given $T$. 
It then 
 changes 
  slope and again follows a 
  $logV_{ac}$ linear relation 
   for each 
   temperature studied between 50 mK and 300 mK. The slope of the $logV_{ac}$ linear relation for each temperature was plotted versus a variety of temperature functions and 
an  interesting relation was found for 
the slope as well as the turning point for $NLRS$ as seen in the plot  
 in Fig.~\ref{Kubota_SS_4}. It was found 
  that the initial slope varies linearly with $1/T^2$ dependence and then it makes a jump and continues to follow another linear relation, whereas the $NLRS_{turning}$ point value is also found to follow a linear relation over some $1/T^2$ range and then jumps to another linear relation. The jump is happening at $1/T^2 = 0.00025\pm0.0004 K^2$, or $T$ = 59 -69 mK, which coincides with the $T_c$ within the error.

Actually the $logV_{ac}$ linear dependence may be expected, as originally pointed out by Anderson\cite{VFAnderson}, if a tangled quantized vortex state is to be polarized by the excitation or to form linear vortex lines so this result 
 may be related to the polarizability change in the VF state at $T_c$ in addition to the hysteretic component appearance discussed in section 3.1. 
There remains a question as to the origin of the peculiar $1/T^2$ dependence. 
Anderson\cite{anderson2011} pointed out that it may be related to a quantum phase transition, but it needs further study.

\section{Experimental evidence for the quantized vortex lines penetration into the SS 
 state 
 under DC rotation}
\label{sec:4}

Straight quantized vortex lines are introduced into a macroscopic coherent state of a superfluid in a cylindrical vessel under DC rotation above a certain angular velocity $\Omega_{c1}$\cite{hessfairbank}, and at large enough $\Omega >> \Omega_{c1}$, the number of vortices would be changing in proportion to the angular velocity $\Omega$, $n_v$ $\varpropto$ $\Omega$ until it comes close to $\Omega_{c2}$,
 which  is usually far beyond experimentally achievable speed of rotation 
 in the roughly centimeter sized vessels used for the bulk $^4$He.
  One needs a much larger size for the vortex core than 10 $\mu$m to reach the condition beyond $\Omega_{c2}$\cite{4rev/s}. The state beyond $\Omega_{c2}$ means that vortex cores  overlap each other and destroy the superfluidity of the dimensionality defined for the vortex core. It may not necessarily destroy all the coherence of the system, but that of the dimension to which the vortex core is connected. In the present discussion of rotation in a 3D system, DC rotation beyond $\Omega_{c2}(3)$ would destroy the 3D superfluid state. This consideration becomes important when various dimensional subsystems are involved in a single superfluid system. For example, 
   the 2D vortex core diameter is found to be 2.5 nm for the monolayer film superfluid\cite{Shirahama90} and 
there is   an independent 3D vortex core size which is found to be determined by the  controlled pore size of the  porous glass substrates\cite{guzai1}. 
%
%
%
%
%
%
%
%
It had been 
 proposed to detect vortex lines through the SS 
 state of solid $^4$He by TO technique, observing the fact that solid $^4$He 
  also shows evidence of thermal excitations, namely the dissipation peak, presumably vortex rings excitations\cite{VLproposal}. 


An experimental tool to realize vortex lines penetration is rotation of the vessel.  
The ISSP high speed rotation cryostat(ISSP-HSRC), with which 
 vortex penetration as well as rejection phenomena are being studied in the SS 
 state of hcp $^4$He, is described in Yagi $\it et$ $\it al.$\cite{rotcryostat2011}. 
It 
 also reviews how vortex lines in different superfluids have been observed and discusses 
  features of the ISSP-HSRC in detail. 

Since the first report by Kim and Chan, one of the most specific features of the solid He TO responses is 
 the sensitivity to minute AC speed, 
 $V_{ac}$ on the order of 10$\mu$m/s\cite{kimchan1,kimchan2}.
It is the linear edge speed 
 which seems to be important and not amplitude or acceleration according to Aoki, Graves, and Kojima\cite{hysKojima}.
The sensitivity to the linear velocity of the TO responses is also reproduced in our measurements(Fig.~\ref{onset}, Fig.~\ref{VFSS2}, and so on).  Based on 
 these observations it is essential to minimize 
the variations 
 in the rotational speed, in addition to the direct vibrations of the TO itself, for the successful TO study of solid $^4$He. This is easily seen by taking a 10 mm inner diameter torsion bob in which sample He is contained and DC rotational speed of 1 radian/sec with 10$^{-3}$ stability. 
 This situation alone would  already cause variations 
   on the order of 6.3$\mu$m/s 
linear velocity in addition to the usual $V_{ac}$. 
This size of AC velocity change would already cause significant modification of the TO responses. 
One needs to be concerned about actual stability of the rotational speed apart from the average drive velocity accuracy. If the sample diameter were 
larger then the linear velocity effect would be further enlarged and 
would easily cause misleading 
TO results under DC rotation.
%
There is some idea of the constancy of  
ISSP rotating cryostats DC angular speed  
since they have been performing both dilution refrigerator temperature TO experiments on solid He and nuclear demagnetization $T$ 
 range experiments on superfluid $^3$He\cite{ThreeTextures,TextureVortices}. They record  heat leak under DC rotation on the order of 10 nW for the superfluid $^3$He experiments under world record rotational speeds up to 12 radian/sec 
with our first rotating frame.\cite{rotcryo95,rotcryostat2003} and cryostat\cite{rotcryostat2003}. 

\begin{figure}
 \center
  \includegraphics[width=0.5\textwidth]{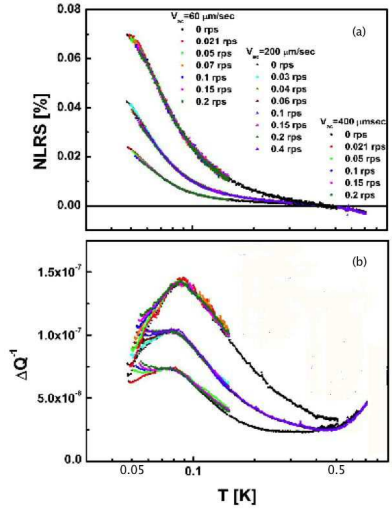}
\caption{TO experiments under DC rotation have been performed in the Kubota group for some time\cite{VLinesThru,VdynamicsUrotation98,rot94,rotcryo95,rotcryostat2003,rotcryostat2011}. $NLRS$(upper frame (a)) is plotted against $T$ for three AC drive linear velocities at the cell edge under DC rotational velocities indicated in revolution/sec (rps). Details of the actual experiments are described 
 elsewhere\cite{SSVLpenetration}. The SS 
  density or $NLRS$ does not change under DC rotation under the present experimental conditions. 
Evidence of the vortex lines penetration through the sample is given (lower frame (b)) by the change in the energy dissipation under DC rotation at constant velocities as given in the figure below a certain temperature, actually below ~75 mK, exactly below $T_c$, derived from the start of hysteretic behavior
\cite{VFtoSS,VFtoSS2011}. Detailed $\Omega$ as well as $T$ dependence analysis will be discussed in the text.
}
\label{TOuRotA}       
\end{figure}

\subsection{Vortex lines penetration experiments in hcp $^4$He}
\label{sec:4.1}

With a specially designed rotating cryostat which does not have a 1K pot for the purpose of condensation of circulating $^3$He 
 for the dilution refrigerator, 
  experiments under DC rotation have been performed for quite some time. 

\begin{figure}
 \center
  \includegraphics[width=0.65\textwidth]
  {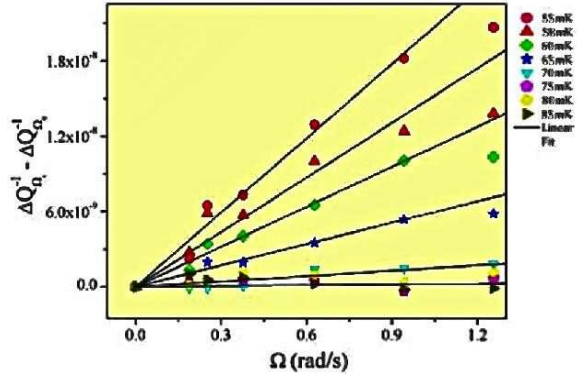}
\caption{$\Omega$ dependence of the 
 change $\Delta$$Q^{-1}_{\Omega}$ - $\Delta$$Q^{-1}_{\Omega=0}$ under DC rotation for the given 
  $T$s. 
 $\Omega$ linear dependence is observed for each $T$ until $\thickapprox$ 0.9 rad/s rotational speed. 
  The $T$ dependence of the slope in this figure may represent 
   $T$ dependence of the SS 
 density. See text below and the next section. Further detailed analysis according to our experiences\cite{VLinesThru,VLproposal} is in progress as well.}
\label{Omega dependence}       
\end{figure}

Let us consider how vortex lines penetrating 
a superfluid can be studied experimentally.  
We proposed a  method for solid $^4$He to detect the vortex lines penetration events\cite{VLproposal} by TO technique, detecting 
extra energy dissipation caused by superflow around the vortex core and the interaction with thermal excitations in the system, similar to the experimental method used for the 3D He film system\cite{VLinesThru}. 
%
%
From 
 the first evidence for the 
penetration by vortex lines as 
presented  at the workshop  in Trieste$\cite{supersolid2008b}$, 
Fig.'s~$\ref{TOuRotA}(a),(b)$ show $NLRS$ and $\Delta$$Q^{-1}$ under DC rotation with given rotational speed for three different 
$V_{ac}$'s.   
While $NLRS$ follows the already discussed $V_{ac}$ dependence\cite{VFPenzev}  it does not change as DC rotational speed changes within the experimental error (see Fig.~$\ref{TOuRotA}(a))$,  
$\Delta$$Q^{-1}$ data in  Fig.~$\ref{TOuRotA}$(b) show a significant change below $T$$\sim$75 mK according to the DC angular velocity  $\Omega$ change for each of  three different AC excitation velocities $V_{ac}$'s. 
The  $\Omega$ dependent change in the $\Delta$$Q^{-1}$ does not seem to depend so much on  $V_{ac}$. 
Fig.~\ref{Omega dependence} shows the almost  linear dependence of $\Delta$$Q^{-1}$ for each of  temperature $T$. 
This behavior certainly 
 supports
  the picture of the vortex lines 
in 
 the SS 
 state of hcp $^4$He. Let us check if this is right by comparing the expected $T$ dependence of such extra energy dissipation.  
  Following the same analysis as for the 3D He film superfluid\cite{VLinesThru}, the expected extra energy dissipation caused by the interaction between the superflow around each of the vortex lines and the thermal excitations in the system would be expressed as follows:

 \qquad  \qquad $\Delta$$Q^{-1}_{\Omega}$   $\varpropto$  $n_v$ $\centerdot$ $\rho_{ss}$($T$) $\varpropto$  $\Omega$ $\centerdot$ $\rho_{ss}$($T$)   \qquad  \qquad    \qquad   (4.1) \\
 \\
%
%
%
%
%
%
where $\Delta$$Q^{-1}_{\Omega}$ is the extra energy dissipation under DC rotation, $n_v$ is the number of vortex lines penetrating the sample, and $\rho_{ss}$($T$) is the SS 
 density, which represents the macroscopic coherence over the whole sample as a function of $T$. Therefore the simplest idea of its $T$ dependence comes from  that of 
$\rho_{ss}$($T$).

\subsection{SS 
 Density $\rho_{ss}$($T$)}
\label{sec:4.2}

\begin{figure}
 \center
  \includegraphics[width=1.0
  \textwidth]{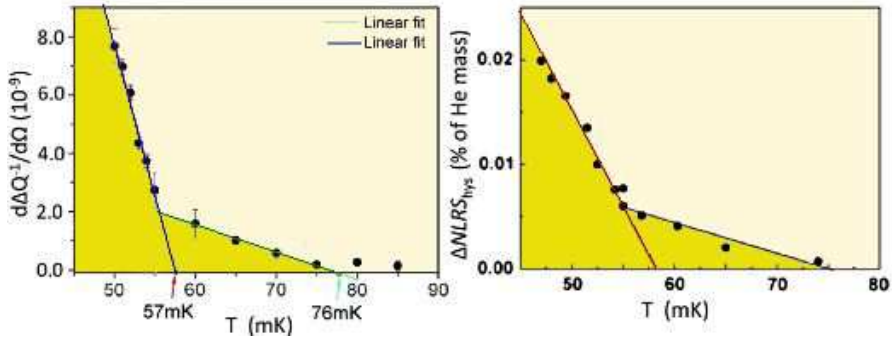}
\caption{ $T$ dependence of the $\Omega$ linear dependent dissipation change under DC rotation for given  $T$'s, compared with the $T$ dependence of the $\rho_{ss}$ obtained by the hysteretic component.}
\label{rho_ss_DeltaQ-1_rot}       
\end{figure}

It is discussed briefly how 
 the value of 
 the SS 
  density $\rho_{ss}$($T$) could be derived from at least two independent experimental quantities,  $d$$\Delta$$Q^{-1}_{\Omega}$/$d\Omega$  and $NLRS_{hys}$ in 4.1. 
Now, 
 how does 
this appear as 
evidence for the real supersolidity in hcp solid $^4$He?  Fig.~\ref{rho_ss_DeltaQ-1_rot}(a) on the left  indicates the slope $d$$\Delta$$Q^{-1}_{\Omega}$/$d\Omega$ of Fig.~\ref{Omega dependence}  as a function of  $T$, whereas (b) on the right 
 the hysteretic component $NLRS_hys$ discussed in Chapter 3 is displayed. 
The former is obtained from the data under DC rotation and the other is data obtained  
without 
rotation. 
 A significant resemblance between them is observed, namely, both quantities appear below the same temperature near 75 mK and furthermore, linear $T$ dependence below 75 mK continues until some temperature 
slightly below 57 mK and then increases  towards lower $T$. 

It is recalled that the hysteretic component of $NLRS$ has shown a critical behavior as seen 
 in Fig.~\ref{VFSS5}. The critical behavior of the SS 
 density with the critical exponent 2/3 with $T_c$= 56.7 mK,  consistent with 3D XY model,  suggests 
  a second order phase transition on one hand, but the existence of the hysteresis implies a first order phase transition. 
 Shimizu$\it et$ $\it {al.}$\cite{VFtoSS}   have discussed a possible weak first order phase transition as the type of 
   transition from the VF 
    into the SS 
  state\cite{VFtoSS2011} with critical behavior. As  has been seen in section 4.2 there is 
  a hint of this kind of transition also in the $log$$V_{ac}$ linear dependence jump in the slope as a property of the 
   VF state 
   as 
    in Fig.~\ref{Kubota_SS_3}, \ref{Kubota_SS_4}. All these 
    observations tell us that the transition between the VF 
    and the SS states is a unique  phase transition. Further 
     study is needed.

Please note, in Fig.~\ref{rho_ss_DeltaQ-1_rot}, the left hand extra energy dissipation divided by $\Omega$ data are taken in the $"$equilibrium$"$ state under DC rotation, whereas the right hand data of the hysteresis component of $NLRS$ were observed in still condition in an additive manner to the $NLRS$ data from higher $T$. What does this result mean? 
The $"$excited$"$ state is discussed as expelling 
 the external vortices in section 4.1 from the upper section of Fig.~\ref{VFSS3}, where the hysteretic energy dissipation change $\Delta$$Q^{-1}_{hys}$ showed a negative change across the hysteretic path difference between the$"$excited$"$ and $"$equilibrium$"$ states.  The result expressed on the right side 
 of Fig.~\ref{rho_ss_DeltaQ-1_rot} indicates that the $"$excited$"$ state is really a Landau-like state, where external vortices are expelled. The 
 $\rho_{ss}$ is an additive quantity to the VF state $NLRS$. 

\subsection{The excited state and the equilibrium state}
\label{sec:4.3}

It has been seen that AC vortices are 
 expelled from the $"$excited$"$ state of  hcp $^4$He which 
 shows Landau-like state behavior. 
  The vortex lines have been observed to penetrate 
 the $"$equilibrium$"$ state sample under DC rotation. All these experimental facts show us that the 
  $\rho_{ss}$  is present even in the $"$equilibrium$"$ state, while the Landau-like state is excited in the hysteretic process below $T_c$.
Then there remains a question 
whether there is a real $\Omega_{c1}$ below which one should expect the real Landau state. Such a question should be answered by future experiments.  
One may be exciting some shielding current for the  $"$excited$"$ state and the $NLRS$ is increased at the same time as external vortices are expelled. 

\section{Supporting experimental results by other groups}
\label{sec:5}

In this chapter we discuss the consistency of our results and their relation 
 to the results by other groups.
The following subjects are discussed, namely, the pressure dependence of the $NLRS$ extrapolated to $T$ = 0 K, the specific heat anomaly observed around 75 to 100 mK by Chan's group
and the superflow experiments by Ray and Hallock 
next, and then TO experiments on "ultra pure" or single crystal samples with "onset"  near 60 to 79 mK.

%
%
%
%
\textbf{Pressure dependence of $NLRS$ as $T$  $\to 0K$}:
\label{sec:5.1}
One of the most important properties to check consistency with results by other groups is the absolute size of the $NLRS$ extrapolated to $T$ = 0 K $NLRS_0$, since Rittner and Reppy\cite{hysRep} reported a huge change of $NLRS_0$ according to the sample cell geometry, namely to the ratio between surface area and the volume of the solid $^4$He sample. The present author thinks that their findings 
might include some effects of  viscoelasticity. In any case it is important to quote some characteristic property to identify the sample studied. 
Kubota group data, except for the earliest ones in a cell with slightly complicated geometry\cite{anneal}, all show rather low  $NLRS_0$  and are stable during repeated measurements over a period exceeding a year,  as  in Penzev $\it {et}$ $\it {al.}$\cite{VFPenzev}.

\textbf{Specific heat anomaly at 75 -100 mK}:
\label{sec:5.2}
A phase transition of second kind with critical fluctuations would cause a sharp specific heat peak and that of first kind would produce a jump in the specific heat.  There have been 
serious searches by Chan's group and they reported a definite specific heat peak at $T$$\approx$75-100 mK, independent of a small 
$^3$He content\cite{Cpeak07,Cpeak09}.  
One does not yet know the detailed connection to the transition from the VF 
 state into the 
 SS state, but the 
 observation of the transition into SS state from the VF state is consistent with the specific heat peak reported by Lin $\it {et}$ $\it {al.}$\cite{Cpeak07,Cpeak09}.

\textbf{Super flow experiments}:
\label{sec:5.3}
While earlier attempts to observe DC super flow failed\cite{noDCflow1,noDCflow2}, recent flow experiments utilizing the thermomechanical effect with liquid in porous  Vycor to drive the bulk solid He mass flow by Ray and Hallock\cite{massflux} seem to have some correspondence to our observation of the VF and the transition into the SS states\cite{VFtoSS,QFS2009Kubota,VFtoSS2011}. Actually they claim to have observed mass flow below $\sim$600 mK and some drastic change below  $\sim$70 mK and increase of the mobility below this $T$.  Theoretical consideration by Aleinikava, Dedits, and Kuklov\cite{glidesuperclimb} describes a microscopic picture of glide and superclimb of dislocations to explain the above experiment.

\textbf{Ultra pure, single crystal samples experiments}:
\label{sec:5.4}
Another interesting observation, which was made for solid $^4$He samples near 60mK, is  reported by  Clark, West and Chan\cite{NCRI}. They have grown large crystals  with high purity and compared them to  standard, pure $^4$He samples with various qualities. The onset temperature for a number of samples with constant temperature or constant pressure crystallization collapse onto a single value of $\sim$79$\pm$5 mK and the low temperature part is well fitted with what they call the $NCRI$, which is proportional to (1- $T$/$T_c$)$^{\gamma}$, where $\gamma$ $=$ 2/3 with $T_c$ $=$ 0.06 K.    It looks as if the purest (single crystal) samples with smallest amount of imperfections would not have the VF contribution, but only the SS part. The absolute size of their $NLRS$ beyond 0.2 $\%$ is still much larger than our observation for $NLRS$ or $\rho_{ss}$(see Fig.~\ref{VFSS4}).
Nobody yet knows the relation between their results and  the VF to SS transition. 
 It is interesting to see a 3D XY critical behavior for their purest samples. It is important to ask the relative size of the VF state and the SS state contributions to the observable $NLRS$.

\section{Summary and Discussion}
 Results for 
 the VF 
  state and 
  the SS 
 state in hcp $^4$He are summarized as follows:
\\
1]. The VF state was found below $T_o$$\sim$ 500mK. The VF state is characterized by the dynamics of the vortex tangle and its cessation at 30mK.\\
2].The transition from the VF state into the SS state is marked 
 by a) appearance of 
hysteresis below $T_c$ with evolution of the equilibrium state and the excited state, b) the VF state $logV_{ac}$ linear dependence change in the equilibrium state at $T_c$.\\
3]. Vortex lines penetration in the equilibrium state under DC rotation has been detected by the extra energy dissipation signal under rotation. The extra dissipation was found to follow the expected relation (4.1) and its temperature dependence was found to coincide with the SS 
 density $\rho_{ss}$($T$), which was derived from the proposed quantity of the hysteretic component of the $NLRS$, $NLRS_{hys}$.\\
4]. The excited state was discussed to be rejecting external vortices totally and this would lead to a consistent picture of  a Landau-like state with 3].\\
5]. The critical velocity $V_c$($T$=0) is evaluated to be about 1 cm/s from the $logV_{ac}$ linear dependence of the hysteretic component $NLRS_{ac}$ to the extrapolated value for   $NLRS_{ac}$=$\rho_{ss}$=0. This is in reasonable agreement 
 with the evaluation $V_c$=$h$/($m_4$$\xi_0$$\pi$) = 6-12 cm/s for $\xi_0$=25-50nm.\\
%

 The  evidence of the VF state and its transition into the SS state has been shown. It requires some different mechanism than the BCS scenario. 
 The VF triggering SS state appearance is happening, where the VF state is describable by a tangled vortex state model probably involving some thermally excited low dimensional subsystems.
The macroscopic coherence appearance at low $T$ well below $T_c$ should be related to the problem of the ground state discussed by Anderson\cite{groundstateBHmodel}. 
%
%
The shear modulus observation\cite{ShearModulus,statistics} problem has been so far neglected, but it would be essential to construct a real microscopic description of the supersolidity and the vortex state in hcp $^4$He.
So far 
the vortex physics in solid $^4$He has 
 been discussed in 
 the VF state and the SS state, in one of the simplest  Bose quantum crystals. Before solid $^4$He the vortex state had been discussed only 
among so called 
 $"$new type of superconductors$"$\cite{QL}. Then, one may ask a question: 
is hcp $^4$He a high temperature superfluid? It would be a fundamental question 
of the present day understanding  of superfluidity,  superconductivity and the vortex state, related to the next question. 
%
Are there further new supersolids awaiting discovery? 
After studying the case of hcp $^4$He, one may ask oneself such a question. 
It would be a new challenge to find one in the second century following 
the first production of a 
superfluid about a century 
 ago.

\begin{acknowledgements}
Our own developments with higher stability TO experiments under rotation involve decades 
 of the following people's efforts after the initial TO cooperation involving Takashi Watanabe, Nobuo Wada, and Keiya Shirahama: Takeshi Igarashi,  Goh Ueno, Takeshi Iida, Vladimir Kovacik, Maxim Zalalutdinov, Muneyuki Fukuda, Toshiaki Obata,Vitaly Syvokon, Nikoley Mikhin, Itsumi Tanuma, Yuji Ito, Ken Izumina, Andrei Penzev, Yoshinori Yasuta, Nobutaka Shimizu,  Akira Kitamura, and Masahiko Yagi. 
 Contributions are appreciated by other members of Kubota group in the development and technical support by Hisahiro Hamada of Kohzu Seiki co., Masahito Sawano, Kiyoshi Akiyama of Rigaku co.   
Cooperation with J.D. Reppy, and Kris Rogacki and theoretical support by  Edoard Sonin, Tomoki Minoguchi, and Sergey Nemirovskii has been indispensable.
The author thanks Robert Mueller for the continued encouragement and for correction in the author's English expressions over the whole period. 
\end{acknowledgements}
$\\$


\bibliographystyle{unsrt}
\bibliography{mk2012_02_01}

\begin{thebibliography}{10}

\bibitem{kimchan1}
E.~Kim and M.~H.~W. Chan.
\newblock 
\newblock {\em NATURE}, 427:225--227, 2004.

\bibitem{kimchan2}
E.~Kim and M.~H.~W. Chan.
\newblock 
\newblock {\em science}, 305:1941, 2004.

\bibitem{VFAnderson}
P.~W. Anderson.
\newblock {\em Nature Physics}, 3:160 -- 162, 2007.

\bibitem{VFAnderson2}
P.~W. Anderson.
\newblock {\em Phys. Rev. Lett.}, 100:215301--1--3, 2008.

\bibitem{VFPenzev}
A.~Penzev, Y.~Yasuta, and M.~Kubota.
\newblock {\em Phys. Rev. Lett.}, 101:065301, 2008.

\bibitem{Nemir0907}
S.~K. Nemirovskii, N.~Shimizu, Y.~Yasuta, and M.~Kubota.
\newblock arxiv:0907.0330.

\bibitem{VFtoSS}
N.~Shimizu, Y.~Yasuta, and M.~Kubota.
\newblock arxiv:0903.1326, 2009.

\bibitem{QFS2009Kubota}
M.~Kubota, N.~Shimizu, Y.~Yasuta, P.~Gumann, and S.~K. Nemirovskii.
\newblock {\em J. Low Temp. Phys.}, 158:572 --577, 2010.

\bibitem{VFtoSS2011}
M.~Kubota, N.~Shimizu, Y.~Yasuta, A.~Kitamura, and M.~Yagi.
\newblock {\em J Low Temp Phys}, 162:483--491, 2011.

\bibitem{ReattoChester}
L.~Reatto and G.~V. Chester.
\newblock {\em Phys. Rev.}, 155(1):88 -- 100, 1967.

\bibitem{Chester70}
G.~V. Chester.
\newblock {\em Phys. Rev. A}, 2(1):256--258, 1970.

\bibitem{AndreevLifshitz69}
A.~F. Andreev and I.~M. Lifshitz.
\newblock {\em JETP}, 29:1107--1113, 1969.

\bibitem{Leggett70}
A.~J. Leggett.
\newblock {\em Phys. Rev. Lett.}, 25(22):1543--1546, 1970.

\bibitem{matsudatsuneto70}
H.~Matsuda and T.~Tsuneto.
\newblock {\em Progress of Theoretical Physics}, Supplement,(46):411-- 436,
  1970.

\bibitem{CuprateSC}
J.~G. Bednorz and K.~A. Mueller.
\newblock {\em Z. Phys.}, 64:189, 1986.

\bibitem{SuperFlowDissipationRMP}
P.~W. Anderson.
\newblock {\em Rev. Mod. Phys.}, 38(2):298--310, 1966.

\bibitem{SF_BEC}
F.~London.
\newblock {\em Phys. Rev.}, 34:947--954, 1938.

\bibitem{BCS}
J.~Bardeen, L.~N. Cooper, and J.~R. Schrieffer.
\newblock {\em Phys. Rev.}, 108:1175, 1957.

\bibitem{Feynmann}
Feynmann.
\newblock volume~1 of {\em Progress in Low Temperature Physics}.
\newblock North-Holland, Amsterdam, 1955.

\bibitem{WilliamsLambdaTransition}
Garry~A. Williams.
\newblock {\em Phys. Rev. Lett.}, 59(17):1926 -- 1929, 1987.

\bibitem{VS}
G.~Blatter, M.~V. Feigel'man, V.~B. Geshkenbein, A.~I. Larkin, and V.~M.
  Vinokur.
\newblock {\em Rev. Mod. Phys.}, 64(4):1125--1388, 1994.

\bibitem{FeBasedSC}
Y.~Kamihara and et~al.
\newblock 
\newblock {\em J. Am. Chem. Soc.}, 128:10012--10013, 2006.

\bibitem{FFH}
D.~S. Fisher, M.~P.~A. Fisher, and D.~A. Huse.
\newblock {\em Phys. Rev. B}, 43(1):130, 1991.

\bibitem{clem97}
J.~R. Clem, Thomas Pe, and M.~Benkrauda.
\newblock {\em Physica C}, 282-287:311, 1997.

\bibitem{AndersonSupersolid}
P.~W. Anderson.
\newblock Theory of supersolidity.
\newblock {\em arXiv:1111.1707}, pages 1--15, 2011.

\bibitem{B}
V.~L. Berezinskii.
\newblock {\em Sov. Phys. JETP}, 34:610, 1972.

\bibitem{KT}
J.~M. Kosterlitz and D.~J. Thouless.
\newblock {\em J. Phys.}, c 5:L124, 1972.

\bibitem{UnivJump}
D.~R. Nelson and J.~M. Kosterlitz.
\newblock {\em Phys. Rev. Lett.}, 39(19):1201--1205, 1977.

\bibitem{bishopreppy78}
D.~J. Bishop and J.~D. Reppy.
\newblock {\em Phys. Rev. Lett.}, 40(26):1722--1730, 1978.

\bibitem{bishopreppy80}
D.~Bishop and J.~Reppy.
\newblock {\em Phys. Rev. B}, 22:5171, 1980.

\bibitem{AHNS}
Vinary Ambegaokar, B.~I. Halperin, David~R. Nelson, and Eric~D. Siggia.
\newblock {\em Phys. Rev. B}, 21(5):1806--1825, 1980.

\bibitem{KubotaQVPhys}
Minoru Kubota.
\newblock {\em J Low Temp PhysL}, 2012.

\bibitem{shev87}
S.~I. Shevchenko.
\newblock {\em Sov. J. Low Temp. Phys.}, 13:61, 1987.

\bibitem{shev88}
S.~I. Shevchenko.
\newblock {\em Sov. J. Low Temp. Phys.}, 14:553, 1988.

\bibitem{Prokfev}
Nikolay Prokf'ev.
\newblock {\em Advances in Physics}, 56(2):381--402, 2007.

\bibitem{GalliReattoR}
D.~Galli and L.~Reatto.
\newblock {\em J. Phys. Soc. JPN}, 77(11):111010--1--16, 2008.

\bibitem{SSDisorder}
S.~Balibar and F.~Caupin.
\newblock {\em J. Phys.: Condens. Matter}, 20:173201--1--19, 2008.

\bibitem{Balibar2010}
Sebastien Balibar.
\newblock {\em Nature}, 464:176 -- 182, 2010.

\bibitem{ShearModulus}
James Day and John Beamish.
\newblock {\em Nature}, 450:853 -- 856, 2007.

\bibitem{statistics}
J.~T. West, O.~Syshchenko, J.~Beamish, and M.~H.~W. Chan.
\newblock {\em Nature Physics}, 5:598 -- 601, 2009.

\bibitem{aboveTc}
Yayu Wang, Lu~Li, M.~J. Naughton, G.~D. Gu, S.~Uchida, and N.~P. Ong.
\newblock {\em Phys. Rev. Lett.}, 95:247002--1--4, 2005.

\bibitem{QL}
A.~J. Leggett.
\newblock {\em Quantum Liquids}.
\newblock Oxford University Press, 2006.

\bibitem{highp}
E.~Kim and M.~H.~W. Chan.
\newblock {\em Phys. Rev. Lett.}, 97:115302--1--4, 2006.

\bibitem{ReppyTylar}
J.~D. Reppy and A.~Tylar.
\newblock {\em Excitations in Two Dimensional and Three Dimensional Quantum
  Fluids}, volume 291-300.
\newblock Plenum Press, NewYork, 1991.

\bibitem{FukudaPhD}
M.~Fukuda.
\newblock PhD thesis, Dept. Physics, the university of Tokyo, 1999.

\bibitem{VLinesThru}
M.~Fukuda, M.~K. Zalalutdinov, V.~Kovacik, T.~Minoguchi, T.~Obata, M.~Kubota,
  and E.~B. Sonin.
\newblock {\em Phys. Rev. B}, 71:212502, 2005.

\bibitem{tsubota2004}
M.~Tsubota, C.~Barenghi, T.~Araki, and Mitani A.
\newblock {\em Phys. Rev. B}, 69:134515, 2004.

\bibitem{nemir0902}
S.~K. Nemirovskii.
\newblock {\em Phys. Rev. B}, 81:064512, 2010.

\bibitem{hysKojima}
Y.~Aoki, J.~C. Graves, and H.~Kojima.
\newblock {\em Phys. Rev. Lett.}, 99:015301--1--4, 2007.

\bibitem{hysChan}
A.C. Clark, J.D. Maynard, and M.H.W. Chan.
\newblock {\em Phys. Rev. B}, 77:184513, 2008.

\bibitem{hysRep}
Ann~S. Rittner and John~D. Reppy.
\newblock {\em Phys. Rev. Lett.}, 101:155301--1--4, 2008.

\bibitem{Josephsonslength1}
D.~J. Josephson.
\newblock {\em Phys. Lett.}, 21:608, 1966.

\bibitem{Josephsonslength2}
M.~E. Fisher, M.~N. Barber, and D.~Jasnow.
\newblock {\em Phys. Rev. A}, 8:1111--, 1978.

\bibitem{guzai1}
N.~P. Mikhin, V.~E. Syvokon, T.~Obata, and M.~Kubota.
\newblock {\em Physica B}, 329--333:272--273, 2003.

\bibitem{anderson2011}
P.W. Anderson.
\newblock private communication, June 2011.

\bibitem{hessfairbank}
G.~B. Hess and W.~M. Fairbank.
\newblock {\em Phys. Rev. Lett.}, 19(5):216--218, 1967.

\bibitem{4rev/s}
T.~Obata, I.~Tanuma, T.~Igarashi, and M.~Kubota.
\newblock {\em J. Low Temp. Phys.}, 138(3/4):929--932, 2005.

\bibitem{Shirahama90}
K.~Shirahama, M.~Kubota, S.~Ogawa, N.~Wada, and T.~Watanabe.
\newblock {\em Phys. Rev. Lett.}, 64(13):1541--1544, 1990.

\bibitem{VLproposal}
M.~Kubota, M.~Fukuda, T.~Obata, A.~Ito, A.~Penzev, T.~Minoguchi, and E.~B.
  Sonin.
\newblock {\em AIP Conf Proc}, 850:283--284, 2006.

\bibitem{rotcryostat2011}
M.~Yagi, A.~Kitamura, N.~Shimizu, Y.~Yasuta, and M.~Kubota.
\newblock {\em J Low Temp Phys}, 162:754--761, 2011.

\bibitem{ThreeTextures}
R.~Ishiguro and $\it {et}$~$\it {al.}$.
\newblock {\em Phys. Rev. Lett.}, 93(12):125301--1--4, 2004.

\bibitem{TextureVortices}
M.~Yamashita and $\it {et}$~$\it {al.}$.
\newblock {\em Phys. Rev. Lett.}, 94:075301 --1--4, 2005.

\bibitem{rotcryo95}
V.~Kovacik, M.~Fukuda, T.~Igarashi, T.~Torizuka, M.~K. Zalalutdinov, and
  M.~Kubota.
\newblock {\em J Low Temp Phys}, 101(3/4):567--572, 1995.

\bibitem{rotcryostat2003}
M.~Kubota and et~al.
\newblock {\em Physica B}, 329-333:1577--1581, 2003.

\bibitem{VdynamicsUrotation98}
M.~Fukuda, M.K. Zalalutdinov, V.~Kovacik, T.~Igarashi, T.~Obata, K.~Ooyama, and
  M.~Kubota.
\newblock {\em J Low Temp Phys}, 113(3/4):417 -- 422, 1998.

\bibitem{rot94}
M~Kubota, G~Ueno, T~Igarashi, and Y~Karaki.
\newblock {\em Physica B}, 194-196:797, 1994.

\bibitem{SSVLpenetration}
M.~Yagi, A.~Kitamura, N.~Shimizu, Y.~Yasuta, and M.~Kubota.
\newblock {\em J Low Temp Phys}, 162:492--499, 2011.

\bibitem{supersolid2008b}
Trieste August 2008.

\bibitem{anneal}
A.~Penzev, Y.~Yasuta, and M.~Kubota.
\newblock {\em J Low Temp Phys}, 148:677--681, 2007.

\bibitem{Cpeak07}
X.~Lin, A.C. Clark, and M.H.W. Chan.
\newblock {\em nature}, 449:1025

\bibitem{Cpeak09}
X.~Lin, A.~Clark, Z.~Cheng, and M.H.W. Chan.
\newblock {\em Phys. Rev. Lett.}, 102:125302, 2009.

\bibitem{noDCflow1}
J.~Day, T.~Herman, and J.~Beamish.
\newblock {\em Phys. Rev. Lett.}, 95:035301, 2005.

\bibitem{noDCflow2}
J.~Day and J.~Beamish.
\newblock {\em Phys. Rev. Lett.}, 96:05304, 2006.

\bibitem{massflux}
M.~Ray and R.~Hallock.
\newblock {\em Phys. Rev. Lett.}, pages 145301--1--4, 2010.

\bibitem{glidesuperclimb}
D.~Aleinikava, E.~Dedits, and A.~Kuklov.
\newblock {\em J Low Temp Phys}, 162:464, 2011.

\bibitem{NCRI}
A.C. Clark, J.T. West, and M.H.W. Chan.
\newblock {\em Phys. Rev. Lett.}, 99:135302, 2007.

\bibitem{groundstateBHmodel}
P.~W. Anderson.
\newblock 
\newblock {\em arXiv:1102.4797}, 2011.

\end{thebibliography}




\end{document}